# Observation of Transient Trion Induced by Ultrafast Charge Transfer in Graphene/MoS$_2$ Heterostructure

*Chen Wang[1], Yu Chen[2], Qiushi Ma[3], Peng Suo[1,4*], Kaiwen Sun[1], Yifan Cheng[1], Xian Lin[1], Weimin Liu[2,*], Guohong Ma[1,4*]*

[1]Department of physics, Shanghai University, Shanghai 200444, China

[2]School of Physical Science and Technology, ShanghaiTech University, Shanghai 201210, China

[3]Department of Chemistry, Purdue University, West Lafayette, IN 47907 USA

[4]Institute for quantum science and technology, Shanghai University, Shanghai 200444, China

**ABSTRACT**

Van der Waals (Vdw) heterostructures constructed from TMDCs provide an ideal platform for exploring various quasiparticle behaviors, with trion—composed of neutral exciton and charged carrier—being a notable example. There are typically three methods to generate trion: electrical doping, chemical doping, and direct optical doping. The first two methods generate static trion, while the last gives rise to transient trion. Here, we present an indirect optical doping approach to generate transient trion via ultrafast charge transfer (CT) and achieve control over the trion-to-exciton ratio by adjusting CT in Gr/MoS$_2$ heterostructure. Furthermore, we demonstrated that dynamics of the transient trion generated with this method, which shows slightly longer lifetime than that of exciton accounted for the Coulomb interactions between trion and charged defect. This study provides fresh perspectives on the construction of new quasiparticles, dynamical characterization and the control of the many-body interaction in two-dimensional structure.

**INTRODUCTION**

As the thickness of transition metal dichalcogenides is reduced to the atomic scale, unique optical and electronic properties emerge due to quantum confinement and reduced Coulomb screening, such as layer-dependent bandgap, large exciton binding energy and strong electron-hole Coulomb interactions, etc.[1-4] In fact, the reduced dielectric screening also increases the interactions between excitons as well as exciton and charged carriers, resulting in more stable many-body complexes that survive up to room temperature.[5-8] Studying these many-body complexes enhances our understanding of quasiparticle physics such as many-body interactions, and lays the groundwork for the development of advanced quantum and high-sensitivity optoelectronic devices.[9-12]

Trion is a common many-body complex in TMDCs, typically formed under specific doping conditions.[13-15] Physically, trion is a lower-energy complex that is composed of an exciton surrounded by a Fermi sea of free charge carriers. With increasing doping levels of charged carriers, trion can further evolve into exciton-polariton.[16] Doping is essential for the formation of trion and typically involves three methods. The first is electrical doping, achieved by fabricating TMDCs field-effect transistor device and tuning the gate voltage. This method is widely used for exploring the evolution of many-body complexes with doping due to its advantage of continuously varying the doping concentration.[17-19] The second is chemical doping, which involves altering the elemental composition during the sample growth process or introducing new impurities. Although this method can produce the sample with good stability, it is challenging to precisely control the doping concentration during fabrication.[20-22] The third method is direct optical doping, which involves generation of free electrons and holes in TMDCs through optical excitation above their bandgap, the concentration of trion is determined by the balance between the exciton and trion, which is dominated by the binding energy of exciton and trion .[23, 24] These three methods are typical for altering the doping level of samples, with the first two methods corresponding to the formation of static trion and the last corresponding to transient trion. Thanks to recent advances in studying interlayer charge transfer in Vdw heterostructures of two-dimensional TMDCs and the electrically tunable Fermi level of graphene, the Vdw heterostructures composed of graphene and TMDCs are expected to become a new platform for creating many-body quasiparticles such as trion.[25, 26] This includes static charge transfer doping across energy barriers and dynamic charge transfer doping where hot electrons in graphene comes across energy barriers. For the latter case, the interface charge relaxation dynamics will inevitably affect the behavior of quasiparticle such as trion.

In this work, we explore the potential of constructing transient trion via indirect optical doping through ultrafast interlayer charge transfer, and investigate the relaxation dynamics of transient trion, using transient and static absorption spectroscopy as well as transient terahertz (THz) spectroscopy. Our findings demonstrate that transient trion can be generated through ultrafast interlayer charge transfer, and that the proportion of trion can be tuned by controlling the amount of charge carrier using graphene as a knob. The relaxation dynamics of transient trion indicates that its lifetime is slightly longer than that of exciton. Considering the role of defect in mediating the backflow of doped electron, we attribute the slightly longer lifetime to the Coulomb repulsion between trion and charged defect. Our study presents a method for generating transient trion in graphene-TMDCs heterostructures using ultrafast interlayer charge transfer, offering new insights into the formation and subsequent dynamics of the related many-body complexes.

**RESULTS AND DISCUSSION**

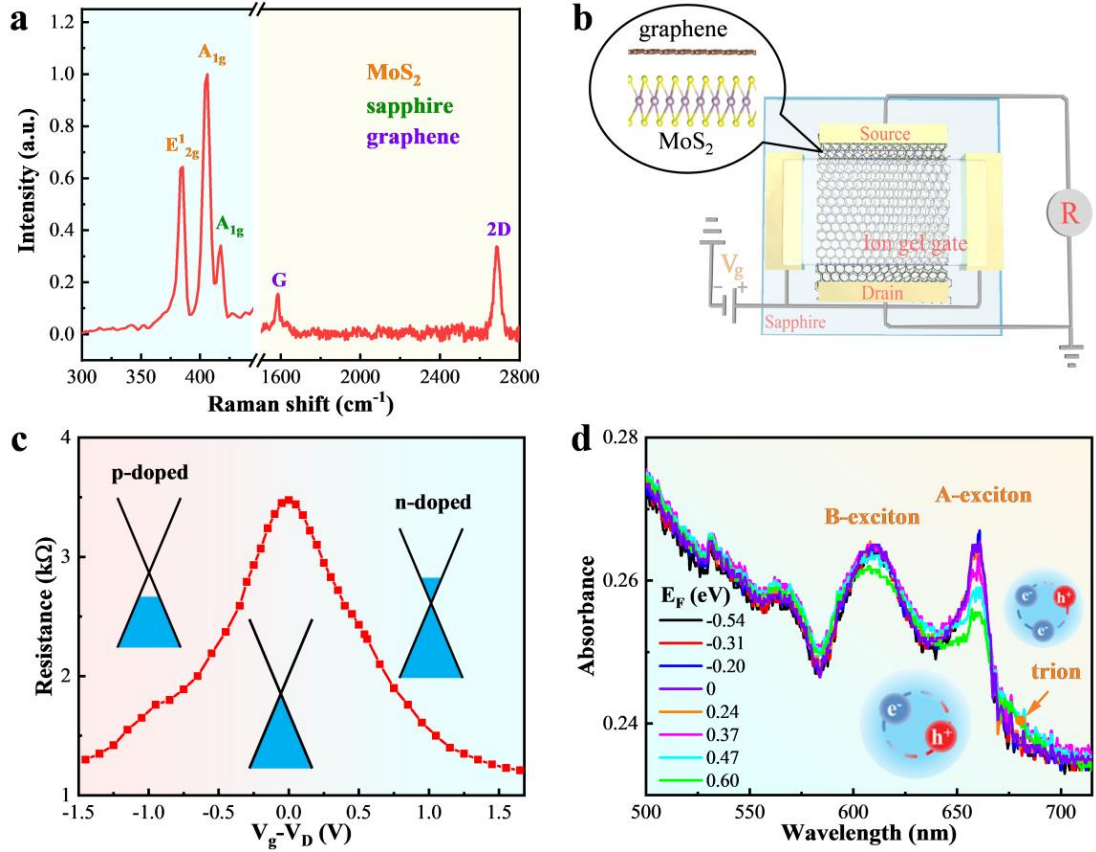

**Figure 1.** Structure and characterization of Gr/MoS$_2$ heterostructure. (a) Raman spectra of Gr/MoS$_2$ heterostructure. (b) Schematic diagram of the Gr/MoS$_2$ heterostructure device. (c) Resistance measurement between the source and drain of Gr/MoS$_2$ heterostructure device under different gate biases. $V_g$ and $V_D$ represent the gate bias beyond and around Dirac point of graphene, respectively. (d) The static absorption spectra of Gr/MoS$_2$ heterostructure devices with different $E_F$ of graphene.

**Figure 1** shows the structure and characteristics of the fabricated field effect transistor (FET) device. The Gr/MoS$_2$ heterostructure on 1-mm sapphire substrate is available commercially (provided by SixCarbon Technology, Shenzhen China), which was fabricated by chemical vapor deposition (CVD) method. **Figure 1a** shows the Raman spectra of the Gr/MoS$_2$ heterostructure. Where the characteristic peaks labeled as A$_{1g}$ and E$^1_{2g}$ represent the out-of-plane and in-plane vibrational peaks of monolayer MoS$_2$, respectively.[27-29] The G-band and 2D-band of graphene are also labeled in the figure.[30, 31] The peak around 418 cm$^{-1}$ (A$_{1g}$) arises from sapphire substrate.[32] It should be noted that the Raman shift were measured before the device being fabricated. **Figure 1b** illustrates the schematic of the Gr/MoS$_2$ heterostructure device. Here, ionic gel is used as the gate, allowing the Fermi level of graphene to be tunable by applying a gate voltage.[33] The detailed fabrication process of the FET device, as well as the relationship between gate voltage and the graphene Fermi level, is provided in S1 of Supporting Information (SI).[34] **Figure 1c** shows the variation in resistance between the source and drain electrodes under different gate voltages. As the graphene Fermi level shifts from the doped state towards the Dirac point, the resistance of graphene increases, indicating that the Fermi level of graphene has been successfully tuned via the gate voltage.[34] **Figure 1d** displays the static absorption spectra of the device under different $E_F$. We observe that as the graphene Fermi level increases, the static absorption spectrum of the device remains unchanged at low gate voltage. However, once the Fermi level of graphene reaches 0.25 eV, the exciton peak absorption begins to decrease, and a new peak appears on the low-energy side of A-exciton (marked as the trion peak), indicating a conversion from exciton to trion. To more clearly illustrate the changes in the optical spectra, the static absorbance change (ΔA) spectra of the device are calculated by subtracting the absorbance at the Dirac point from the absorbance at various Fermi levels, which is shown in **Fig. 2a**.

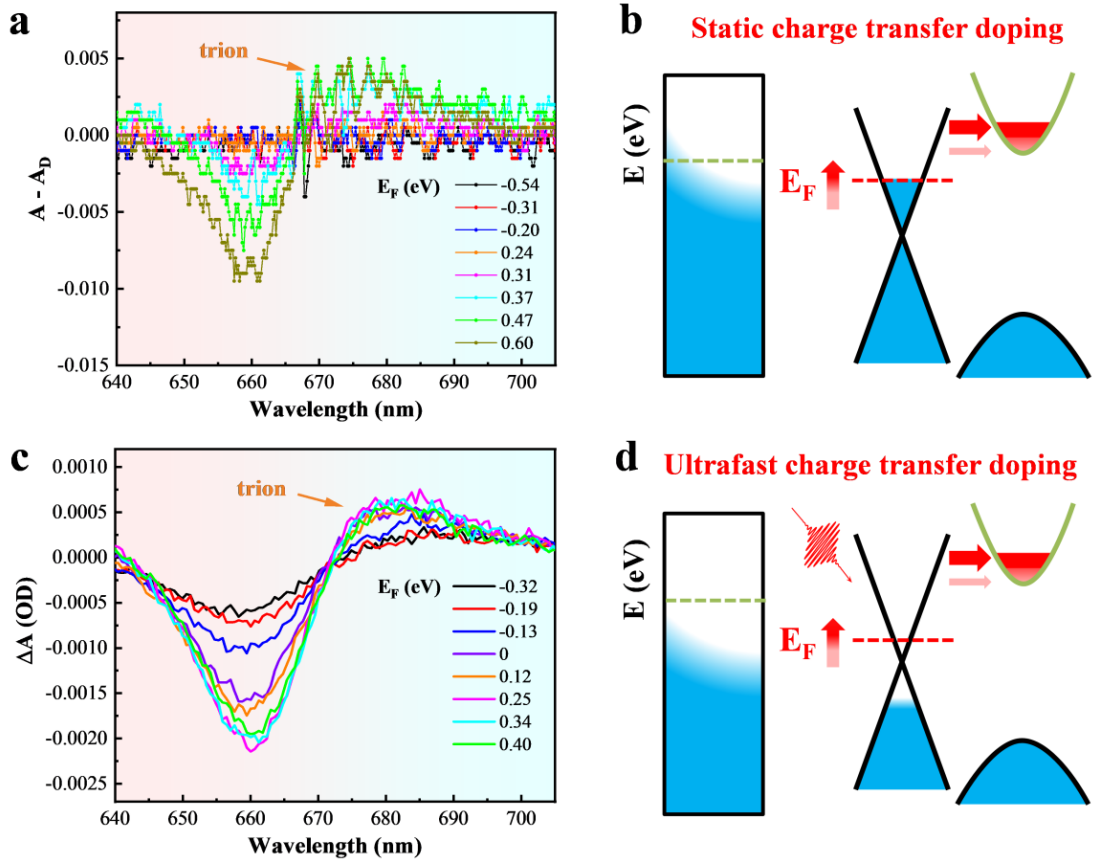

**Figure 2.** Gate voltage dependence of static absorbance change and transient absorption spectra of Gr/MoS$_2$ heterostructure device. (a) Spectra of absorbance changes calculated by subtracting the absorbance at the Dirac point from the absorbance at various E$_F$. (b) Schematic diagram of the static charge transfer doping. (c) Transient absorption spectra of the heterostructure device for various E$_F$ with a delay time of 1 ps following optical excitation with pump fluence of 63.7 µJ/cm$^2$. (d) Schematic diagram of the ultrafast charge transfer doping.

From **Fig. 2a**, it is evident that when E$_F$ is below 0.25 eV, the absorbance change (ΔA=A(E$_F$)-A(E$_F$=0)) of MoS$_2$ remains almost zero within the measured wavelength, once E$_F$ is higher than 0.25 eV, magnitude of ΔA in MoS$_2$ (taking A-exciton as an example) becomes negative, and a new peak with enhanced absorption appears around ~677 nm, we assigned the peak as the trion absorption in monolayer MoS$_2$. The energy of the peak is approximately 47 meV lower than that of A-exciton peak, which is similar with previous reports on trion.[35, 36] The reduction in the exciton absorption and the generation of trion can be understood through the static charge transfer doping (SCTD) schematically in **Fig. 2b**. Due to gate modulation, the chemical potential of graphene at 300 K exceeds the barrier for static charge transfer. As a result, electrons in graphene can transfer to MoS$_2$ under gate bias. This leads to partial occupation of the free carrier states that constitute the excitonic states (i.e., state filling), making these occupied states unavailable for the formation of new excitons.[37] Consequently, this results in the reduction of the exciton absorption. The static charge transfer also leads to electron doping in MoS$_2$, resulting in the formation of trion. We validated the physical process of static charge transfer by calculating the band structure of the Gr/MoS$_2$ heterostructure, the chemical potential of graphene at 300 K, and the electron distribution at different Fermi levels in S2 of SI.[26] To further demonstrate the formation of trion in the heterostructure, we conducted transient absorption study on the device under different Fermi levels. Here, the pump wavelength was set to 780 nm, which is well below the A-exciton energy of MoS$_2$, indicating that only the graphene layer was excited. All transient absorption signals observed in the MoS$_2$ layer originate from hot electron transfer in the graphene layer.[38, 39] The color maps of the transient absorption spectra at different Fermi levels are shown in S3 of SI. The color maps display the photobleaching signal of exciton centered around 660 nm, and enhanced absorption of trion signal centered around 677 nm, which is consistent with previous observation for a direct optical doping experiment.[24] To more clearly demonstrate the changes in transient absorption caused by ultrafast charge transfer, we extracted transient absorption spectra at delay time of 1 ps with several typical Fermi levels, which is displayed in **Fig. 2c**. For all graphene Fermi levels, both exciton and trion signals are observable clearly. As the graphene Fermi level increases, both exciton and trion signals increase, indicating that the amount of transferred charge increases with the Fermi level. Within a photo-thermal emission model, we can calculate the number of transferred carriers as a function of Fermi level, which is presented in S4 of SI. The result shows that the graphene Fermi level significantly affects the number of thermal electrons transferred from graphene by altering the Fermi distribution. Additionally, we observed that experimentally, once the Fermi level reaches 0.25 eV, the number of transferred electrons no longer increases, and the number of transferred carriers begins to decrease with the increase of Fermi level in graphene. This accounts for the transferred carriers surpassing the band edge of $MoS_2$, leading to an increasing barrier for ultrafast charge transfer in the non-equilibrium state, which can be referred to as the gating limit for ultrafast charge transfer. In short, ultrafast charge transfer doping (UCTD) can be illustrated by the schematic in **Fig. 2d**. When graphene is optically excited at room temperature, the increase in electron temperature causes thermal electrons to surpass the transfer barrier. The amount of the thermal electrons exceeding the barrier can transfer to $MoS_2$, leading to electron doping in $MoS_2$ and the subsequent formation of trion. Additionally, as the Fermi level of graphene increases, the number of transferred carriers increases, which enhances the signals of the exciton and trion.

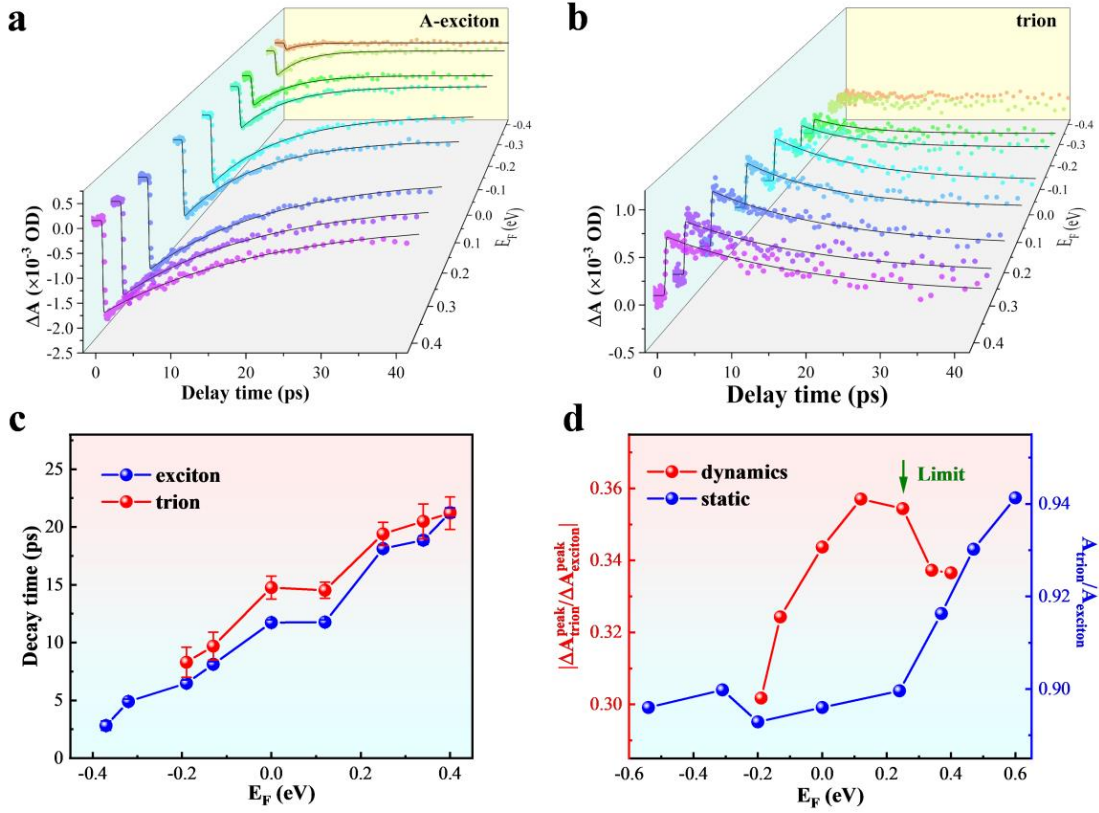

**Figure 3.** $E_F$-dependent lifetimes and magnitude ratios between trion and exciton. (a) Transient absorption spectra of A-exciton for different $E_F$. (b) Transient absorption spectra of trion for different $E_F$. (c) $E_F$-dependent relaxation lifetimes of exciton and trion obtained from single-exponential fits. (d) Magnitude ratio of transient and static absorption peak values for trion and exciton as a function of $E_F$.

Next, we analyze the relaxation dynamics of exciton and trion as well as the ratio of absorbance changes at different $E_F$. We extracted the relaxation curves for A-exciton and trion at their respective wavelengths, as shown in **Fig. 3a** and **3b**. Here, the black solid lines represent the single-exponential fits, and the fitted relaxation lifetimes as a function of $E_F$ are displayed in **Fig. 3c**. We note that with the increase of $E_F$, the lifetimes of both exciton and trion increase, with trion exhibiting a slightly longer lifetime than exciton. Recent studies reveal that backwards electron transfer from $MoS_2$ to graphene occurs after forwards electrons transfer in the graphene/TMDCs heterostructure, and the backwards electron transfer process is mediated by defect states of $MoS_2$. We attribute the longer lifetimes with graphene $E_F$ to the saturation effect of defect states caused by the increasing backflow of transferred electrons.[34, 40] The blue curve in **Fig. 3d** shows the ratio of trion peaking absorbance over that of exciton under static condition, and the red one presents the ratio of peaking absorbance change between the trion and exciton following 780 nm excitation. Obviously, the trion/exciton ratio increases with graphene $E_F$ following optical excitation, this is reasonable that larger amount of transient trion is generated by increasing charge transfer for the higher $E_F$. However, once $E_F$ reaches the gating limit, the decrease in UCTD leads to a reduction in the transient trion proportion. However, the static trion/exciton ratio does not change with $E_F$ when the magnitude of $E_F$ is below gating limit. By contrast, when $E_F$ exceeds the gating limit, this ratio increases with $E_F$ significantly, which accounts for the formation of steady-state trion via SCTD. To summarize, the presence as well as the subsequent dynamics of trion absorption and exciton bleaching following optical excitation, along with the magnitude ratio between trion and exciton, all are associated with the Fermi level of graphene in the heterostructure, suggesting that the proportions of steady-state trion formed by SCTD and transient trion formed by UCTD can be effectively controlled by adjusting the $E_F$ of graphene.

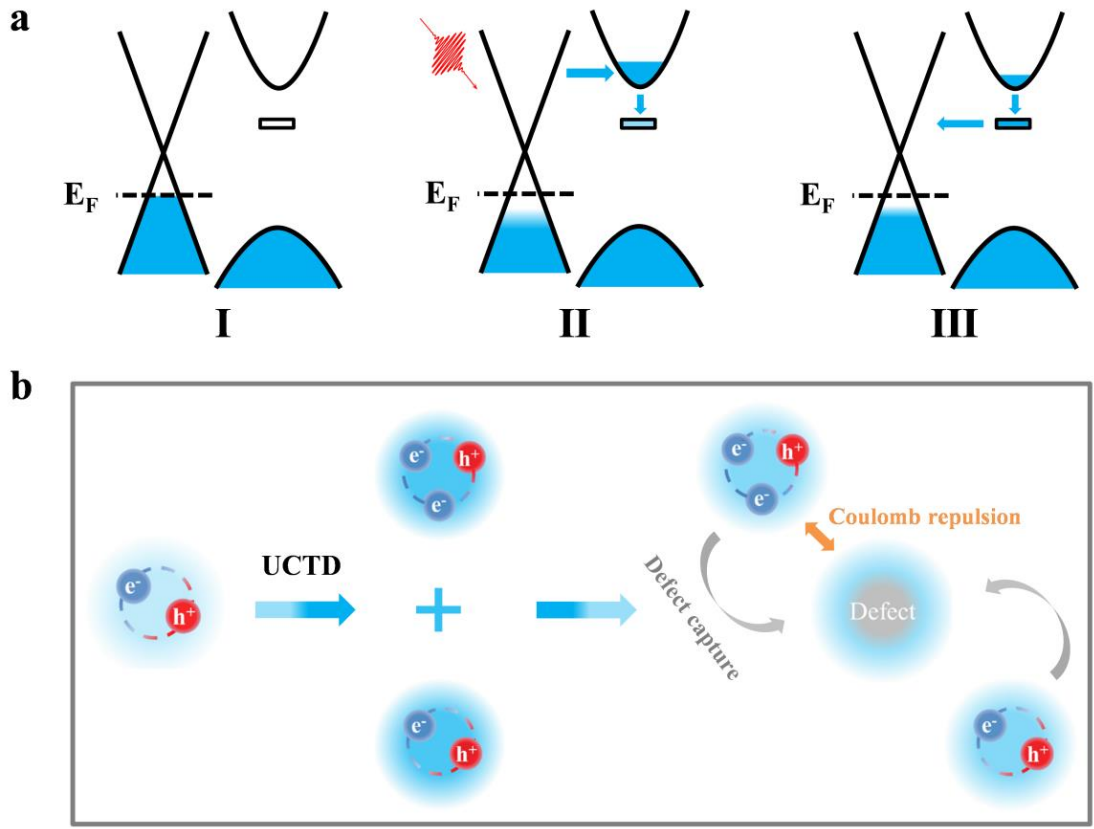

**Figure 4.** Schematic illustration of the defect states-mediated relaxation processes of trion and exciton, as well as electron backflow. (a) Schematic of ultrafast charge transfer and defect states-

mediated electron backflow. (b) Schematic of ultrafast charge transfer doping and defect states-mediated relaxation processes of trion and exciton, where the depth of the blue represents the variation in free carrier concentration caused by charge transfer doping.

Figure 4 summarizes the relaxation processes of the relevant particles in the system, including electron, exciton, and trion. Here, we propose that relaxation processes of these particles are mediated by defect states in $MoS_2$, typically, sulfur vacancies,[41, 42] which includes the backflow of electrons from $MoS_2$, and defect states-assisted non-radiative recombination of exciton and trion. **Figure 4a** illustrates schematically the electron backflow process. Following photoexcitation, ultrafast charge transfer from graphene to $MoS_2$ occurs, and some of the transferred electrons are subsequently trapped by the defect states located below the band edge of $MoS_2$, followed by the backwards electrons transfer from the defect states to graphene. During this process the magnitude of UCTD is expected to decrease gradually as electrons flow back. However, we observed that at longer delay times, the transient trion/exciton ratio does not diminish as expected with the backflow of electrons (see S5 in SI). We consider that this discrepancy might be associated with the relative timescales between electron backflow and exciton. To investigate the electron backflow time, we conducted optical pump and terahertz probe (OPTP) measurements on the device under different gate voltages. Here, the pump wavelength was also set to 780 nm with pump fluence of 47.0 $\mu J/cm^2$, which are presented in S6 of SI. OPTP is very sensitive to the conductivity changes in graphene. We noted that the electron backflow time shows a similar timescale to that of exciton in $MoS_2$, leading us to conclude that the unexpected results are likely due to a combination of the reduction in doping intensity and the decrease in the number of quasiparticles. **Figure 4b** illustrates schematically the relaxation process of quasiparticles. Optical excitation of the heterostructure induces UCTD and increases the free carrier concentration in $MoS_2$, resulting in the simultaneous presence of exciton and trion in $MoS_2$ layer. Subsequently, defect states-mediated non-radiative recombination of exciton and trion takes place. In addition, the lifetime of trion is slightly longer than that of exciton as seen in **Fig. 3c**. We attributed longer lifetime of trion to the Coulomb repulsion between trion and charged defect states (which captures backflow electron). Since exciton is a neutral quasiparticle, and Coulomb repulsion shows negligible influence on exciton. As a result, a little longer lifetime of trion is seen than that of exciton.

**CONCLUSION**

In summary, we utilized both static and transient spectroscopy to demonstrate the potential of constructing transient trion via indirect optical doping enabled by ultrafast interlayer charge transfer in graphene/MoS$_2$ heterostructures. The results indicate that through the ultrafast indirect optical doping method, transient trion can be generated, and its relative proportion to exciton can be controlled by electrically gating the Fermi level of graphene. The relaxation lifetime of trion shows slightly longer than that of exciton, which is attributed to Coulomb repulsion between trion and charged defects. Our study further shows that by modulating Fermi level of graphene, it is possible to construct both steady-state and transient trion within graphene/TMDCs heterostructure simultaneously, providing new insights for the design of related many-body complexes and the development of ultrafast quantum devices.

**Supporting Information**

This PDF file includes: Gr/MoS$_2$ heterostructure device fabrication process and the relationship between gate voltage and graphene Fermi level; Simulation verification of static charge transfer surpassing the barrier; Color maps of the raw data for transient absorption spectra at different Fermi levels; Photo-thermal emission model simulation of $E_F$-dependent ultrafast charge transfer number; Ratio of transient trion to exciton at longer delay times; Optical pump-terahertz probe measurements on Gr/MoS$_2$ heterostructure device under different Fermi levels.

**AUTHOR INFORMATION**

**Corresponding Authors**

**Peng Suo** − Department of Physics, Institute for quantum science and technology, Shanghai University, Shanghai 200444, China; Email: littleboysp@shu.edu.cn

**Weimin Liu** − School of Physical Science and Technology, ShanghaiTech University, Shanghai 201210, China; Email: liuwm@shanghaitech.edu.cn

**Guohong Ma** − Department of Physics, Institute for quantum science and technology, Shanghai University, Shanghai 200444, China; Email: ghma@staff.shu.edu.cn

**Notes**

The authors declare no competing financial interest.

**ACKNOWLEDGMENT**

We acknowledge the financial support from National Natural Science Foundation of China (NSFC, grant NO. 92150101, 12404396); The Postdoctoral Fellowship Program of CPSF (GZB20240418); China Postdoctoral

Science Foundation (2024M751932). Science and Technology Commission of Shanghai Municipality (Grant No. 21JC1402600).

(38) Xu, Z.; Liu, Z.; Zhang, D.; Zhong, Z.; Norris, T. B. Ultrafast dynamics of charge transfer in CVD grown MoS$_2$–graphene heterostructure. *Appl. Phys. Lett.* **2021**, 119.

(39) Tran, M. D.; Lee, S.-G.; Jeon, S.; Kim, S.-T.; Kim, H.; Nguyen, V. L.; Adhikari, S.; Woo, S.; Park, H. C.; Kim, Y.; Kim, J.-H.; Lee, Y. H. Decelerated Hot Carrier Cooling in Graphene via Nondissipative Carrier Injection from MoS$_2$. *ACS Nano* **2020**, 14, 13905-13912.

(40) Krause, R.; Aeschlimann, S.; Chávez-Cervantes, M.; Perea-Causin, R.; Brem, S.; Malic, E.; Forti, S.; Fabbri, F.; Coletti, C.; Gierz, I. Microscopic understanding of ultrafast charge transfer in van der Waals heterostructures. *Phys. Rev. Lett.* **2021**, 127, 276401.

(41) Liu, M.; Shi, J.; Li, Y.; Zhou, X.; Ma, D.; Qi, Y.; Zhang, Y.; Liu, Z. Temperature‐triggered sulfur vacancy evolution in monolayer MoS$_2$/graphene heterostructures. *Small* **2017**, 13, 1602967.

(42) Schuler, B.; Qiu, D. Y.; Refaely-Abramson, S.; Kastl, C.; Chen, C. T.; Barja, S.; Koch, R. J.; Ogletree, D. F.; Aloni, S.; Schwartzberg, A. M. Large spin-orbit splitting of deep in-gap defect states of engineered sulfur vacancies in monolayer WS$_2$. *Phys. Rev. Lett.* **2019**, 123, 076801.
16

**Supporting Information**

Section S1. Gr/MoS$_2$ heterostructure device fabrication process and the relationship between gate voltage and graphene Fermi level.

Section S2. Simulation on static charge transfer surpassing the barrier.

Section S3. Color maps of the raw data for transient absorption spectra at different Fermi levels.

Section S4. Photo-thermal emission model simulation of $E_F$-dependent ultrafast charge transfer number.

Section S5. Ratio between transient trion and exciton at longer delay times.

Section S6. Optical pump-terahertz probe measurements on Gr/MoS$_2$ heterostructure device under different Fermi levels.

**Section S1. Gr/MoS₂ heterostructure device fabrication process and the relationship between gate voltage and graphene Fermi level.**

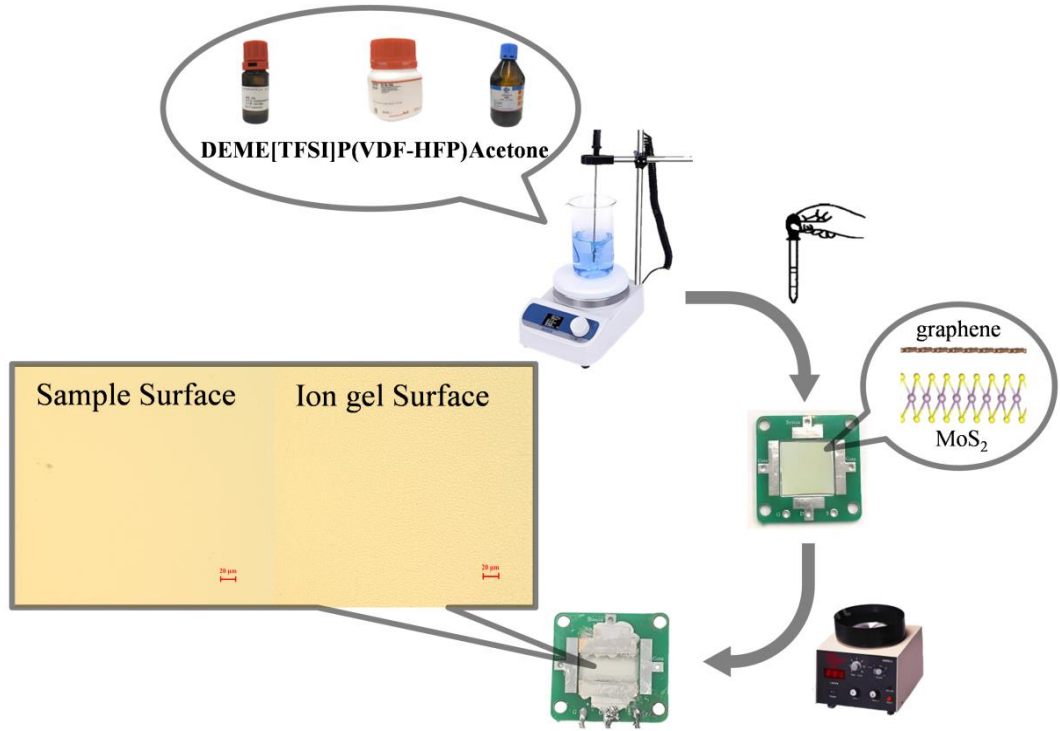

**Figure S1.** Schematic illustration of the fabrication process for the Gr/MoS$_2$ heterostructure device.

Ionic liquid N, N-diethyl-N-methyl-N-(2-methoxyethyl) ammonium bis(trifluoromethylsulfonyl) imide (DEME[TFSI]) and the copolymer poly(vinylidene fluoride-co-hexafluoropropylene), abbreviated as P(VDF-HFP) was used to prepare ion gel. Acetone was used as the solvent. The weight ratio of each component was polymer: ionic liquid: acetone = 1:4:8. The mixture of the three components was stirred for 1 hour to obtain a homogeneous solution of the ion gel, with the temperature of the stirrer set at 75°C. Next, we connected the Gr/MoS$_2$ heterostructure on sapphire substrates to PCB backplanes with different electrodes printed, and covered the source and drain electrodes with adhesive tape for protection. Subsequently, we spin-coated the ion gel onto the top of the heterostructure samples, with the spin coater's speed set at 1000 r/min. Finally, the heterostructure sample spin-coated with ion gel were heated on a 65°C heating stage for 5-10 s, and the adhesive tape protective layer, which lost its adhesion due to heating, was quickly removed. During the process, the ion gel gate electrode spin-coated onto the sample also underwent transparency. **Fig. S1** also shows optical microscopy images of the sample surface and the ion gel gate surface in the device.

The relationship between the gate voltage and the graphene Fermi level is determined by the capacitance of the ion gel gate. We prepared an ion gel layer in a parallel plate capacitor under the same conditions as the fabricated device and measured the capacitance of the ion gel to be 1.5 µF/cm². Based on this, we can calculate the carrier concentration $N$ that can be modulated under different gate biases using Eq. (S1):[1]

$$N = C_{iol\ gel}(V_g - V_D) \tag{S1}$$

Where $V_g$ and $V_D$ are the applied gate bias and the gate bias around the graphene Dirac point, respectively. Subsequently, using the relationship between $E_F$ and the carrier concentration N (Eq. (S2)), we can determine the Fermi level of graphene under different gate biases. Eq. (S2) is:[2]

$$E_F = \hbar v_F \sqrt{\pi N} \tag{S2}$$

Where $\hbar$ is the reduced Planck constant. $v_F = 1.0 \times 10^6$ m/s is the Fermi velocity of graphene.

**Section S2. Simulation on static charge transfer surpassing the barrier.**

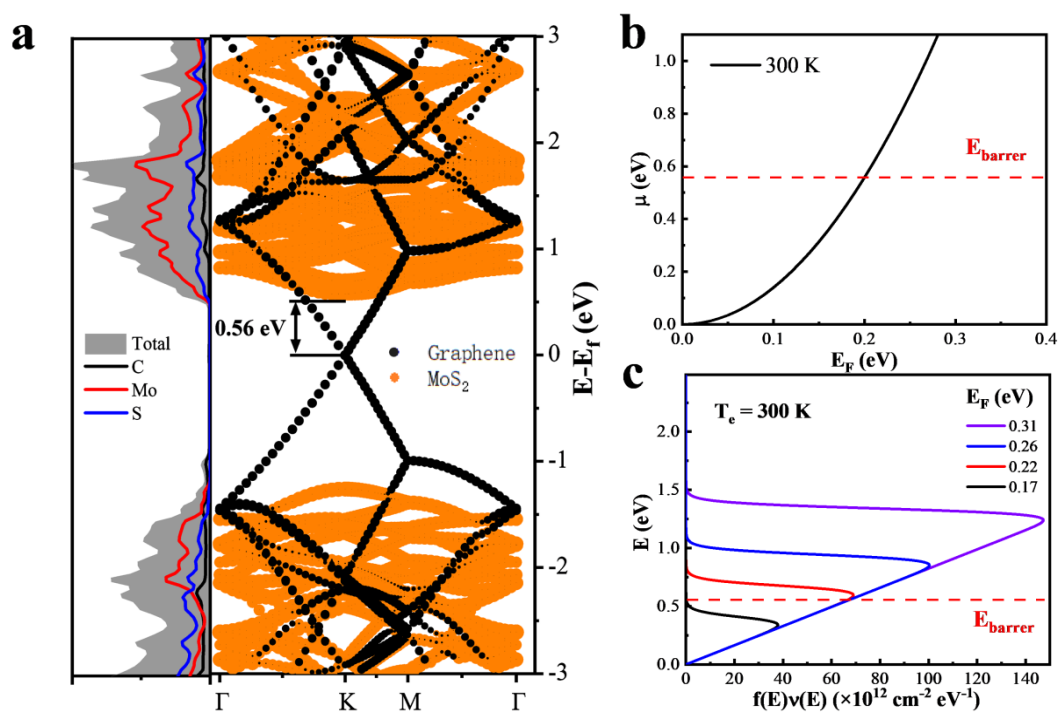

**Figure S2.** Simulation of static charge transfer in the Gr/MoS$_2$ heterostructure achieved by modulating $E_F$. (a) Band structure of the Gr/MoS$_2$ heterostructure obtained from DFT calculations. (b) Chemical potential at different $E_F$ values with an electronic temperature of 300 K (room temperature). (c) Fermi distribution of electrons in graphene at different $E_F$ values at 300 K.

To further validate the static charge transfer occurring at $E_F$ = 0.25 eV in the experiment through simulations, we first calculated the band structure of the Gr/MoS$_2$ heterostructure. The results are shown in **Fig. S2a**. We perform first-principal calculations using density functional theory (DFT) within the Vienna Ab Initio Simulation Package 6.2.1 (VASP 6.2.1).[3] The Gr/MoS$_2$ heterostructure consists of a 4×4 supercell of MoS$_2$ with 48 atoms and a 5 × 5 graphene sheet with 50 atoms. Before constructing the heterostructure, we separately optimize the MoS$_2$ and graphene components. We use the Pedrew-Burke-Ernzerhof (PBE) and the generalized gradient approximation (GGA) methods. The cutoff energy of the plane wave basis set is set to 500 eV. The convergence criteria for electronic energy and ionic forces are set to $10^{-6}$ eV and 0.001 eV/Å, respectively. In the geometry optimization, we use a 3×3×1 Monkhorst-Pack k-point mesh. We calculate the band structures by inserting 20 points between adjacent K-points. After optimization, the interlayer distance between $MoS_2$ and graphene was approximately 3.42 Å. We consider the van der Waals interactions using the Grimme DFT-D3 method. From the calculation results, we obtain the charge transfer barrier $E_{barrier}$ = 0.56 eV. This indicates that static charge transfer can occur once the Fermi distribution of electrons in graphene at 300K exceeds $E_{barrier}$. Next, we calculated the chemical potential and the distribution of electrons in graphene at different $E_F$ at 300K. The expressions for μ is given by Eq. (S3):[4]

$$\mu = \frac{E_F^2}{4 \ln 2 K_B T_e} \quad \text{(S3)}$$

Here, $K_B$ is Boltzmann constant. The equations related to the Fermi-Dirac distribution of electrons are Eq. (S4) and Eq. (S5):

$$f(E) = \frac{1}{e^{(E-\mu)K_B T_e} + 1} \quad \text{(S4)}$$

$$\upsilon(E) = \frac{2E}{\pi \hbar^2 v_F^2} \quad \text{(S5)}$$

Here, μ is the chemical potential, $T_e$ is the electron temperature, ℏ is the reduced Planck constant. From the calculations shown in **Fig. S2b** and **S2c**, we can see that when $E_F$ exceeds 0.20 eV, the chemical potential of graphene at 300K can surpass $E_{barrier}$. This indicates that electrons distributed above $E_{barrier}$ can transfer to $MoS_2$, leading to the occurrence of static charge transfer.

**Section S3. Color maps of the raw data for transient absorption spectra at different Fermi levels.**

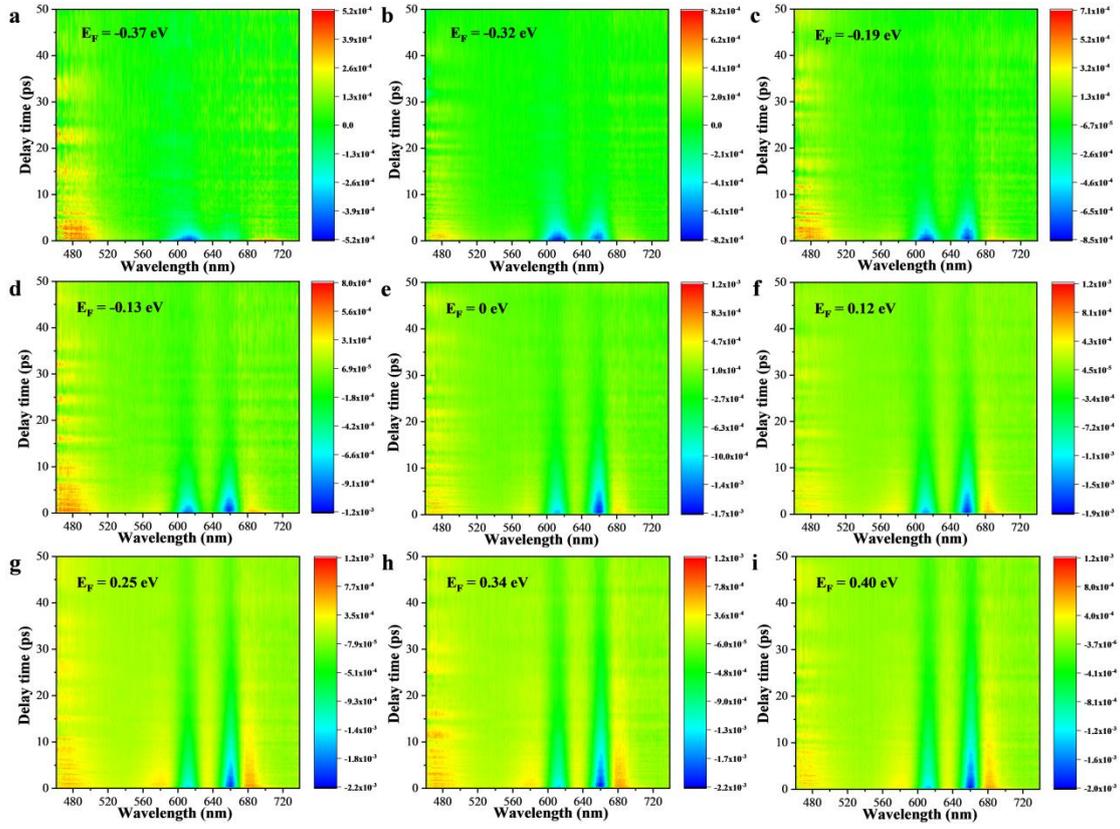

**Figure S3.** The color maps of transient absorption spectra of the device for various $E_F$.

**Section S4. Photo-thermal emission model simulation of $E_F$-dependent ultrafast charge transfer number.**

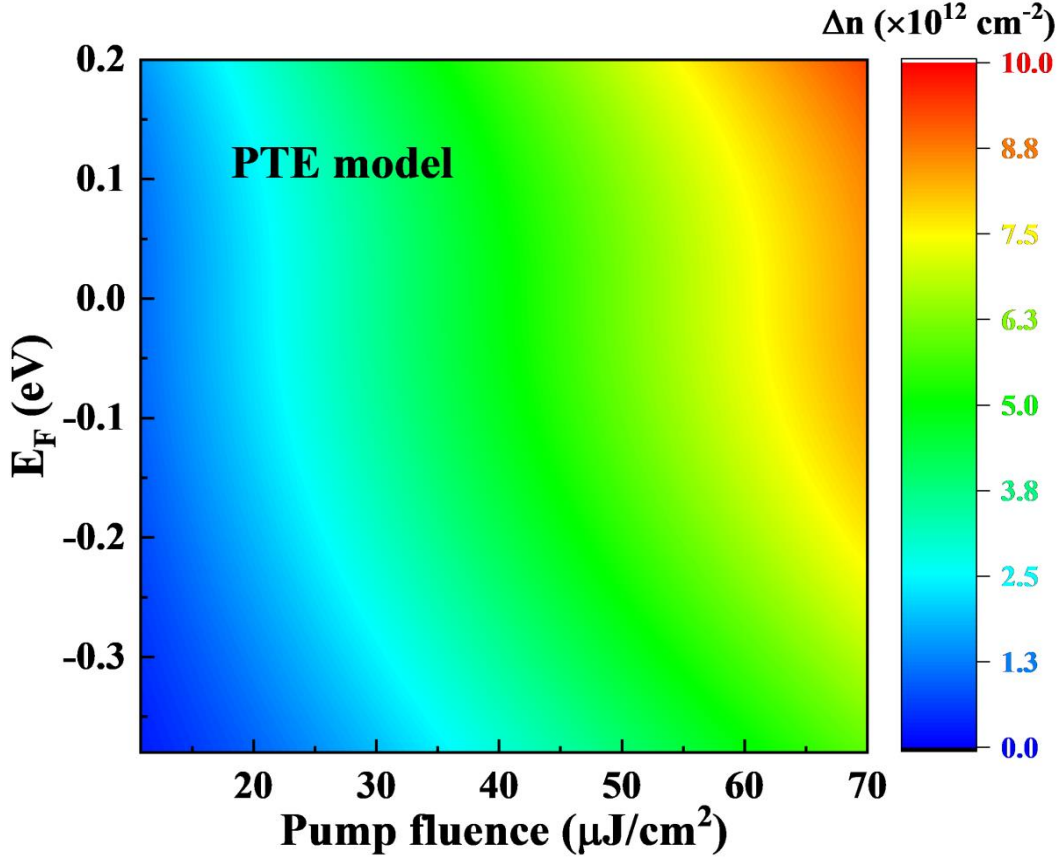

**Figure S4.** Photo-thermal emission model simulation of ultrafast charge transfer number in Gr/MoS$_2$ heterostructure under different $E_F$ and pump fluences.

**Figure S4** shows the Photo-thermal emission model simulation results for ultrafast charge transfer numbers as a function of different $E_F$ and pump fluences. Here, we assume that all electrons in graphene exceeding $E_{barrier}$ are transferred to MoS$_2$, with the calculation described by Eq. (S6):[5]

$$\Delta n = \int_{E_{barrier}}^{\infty} f(E)\, v(E)\, dE \tag{S6}$$

Here, we also need the electron temperature at different pump fluences, which can be obtained using Eq. (S7):

$$T_e = T_L \left(1 + \frac{3\gamma F}{\beta T_L^3}\right)^{\frac{1}{3}} \tag{S7}$$

Here, $T_L$ is the lattice temperature, which is set to be 300 K at room temperature. $\gamma$ is the ratio of the energy absorbed by graphene electron to the incident photon, with $\gamma$ = 70% × 2.3%, according to previous work, 70% of incident photon energy can be applied, whereas graphene absorbs about 2.3% of incident photons. F is the energy density of the incident photons. $\beta = \frac{18\xi(3)K_B^3}{(\pi\hbar v_F)^2}$, $\xi(3) = 1.202$.

**Section S5. Ratio between transient trion and exciton at longer delay times.**

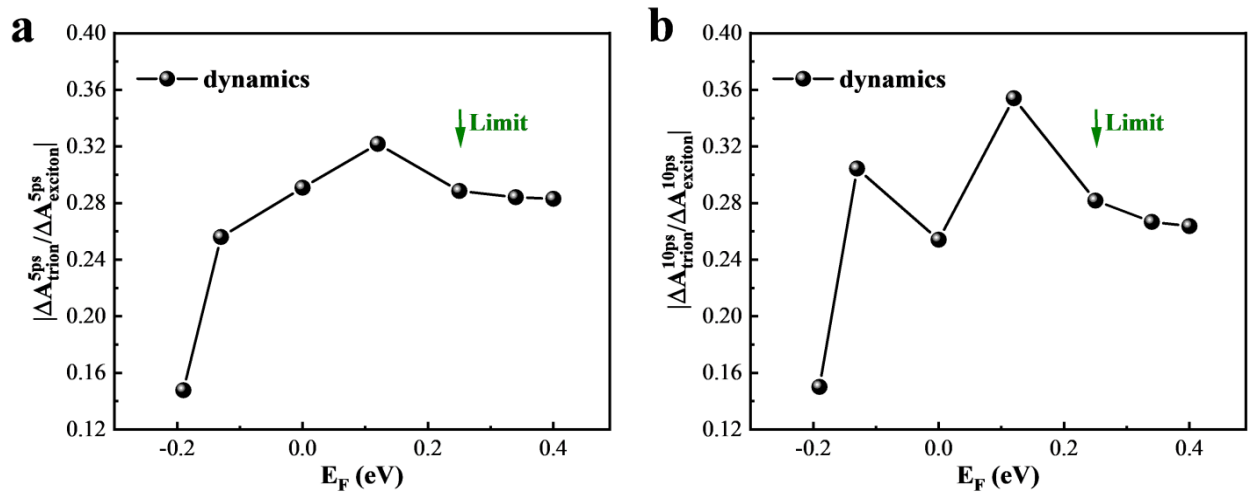

**Figure S5**. Transient trion/exciton radio at different delay times. (a) Trion/exciton at 5 ps delay time. (b) Trion/exciton at 10 ps delay time.

# Section S6. Optical pump-terahertz probe measurements on Gr/MoS$_2$ heterostructure device with different Fermi levels.

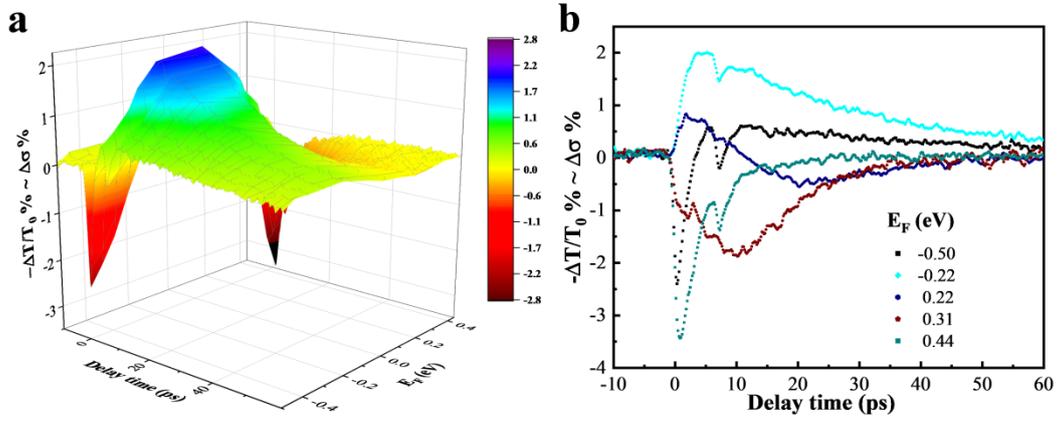

**Figure S6.** Optical pump-terahertz probe(OPTP) measurements on the device. (a)Color map of OPTP measurements under deferent $E_F$. (b) The OPTP test results for several typical $E_F$ extracted from the color map. We observe that the electron backflow time and quasiparticle decay time are similar (tens of picoseconds) for different $E_F$. It is important to note that due to the complexity of the OPTP signals, these refer to the total relaxation times rather than the lifetimes obtained from single-exponential fits in the main text.